\documentclass[11pt]{article}
\usepackage[a4paper,margin=1in]{geometry}
\usepackage{amsmath,amssymb,amsthm}
\usepackage{mathpartir}
\usepackage{listings}
\usepackage{hyperref}

\theoremstyle{plain}
\newtheorem{theorem}{Theorem}[section]

\theoremstyle{definition}

\theoremstyle{remark}
\newtheorem{remark}[theorem]{Remark}

\newcommand{\code}[1]{\mathit{code}(#1)}
\newcommand{\thmT}{\mathbf{th}}
\newcommand{\num}{\mathbf{num}}
\newcommand{\evalU}{\mathbf{U}}
\newcommand{\parse}{\mathbf{parse}}
\newcommand{\stepU}{\mathbf{step}}
\newcommand{\defp}{\mathbf{def}}
\newcommand{\CK}{\mathbf{CK}}
\newcommand{\Kc}{K}
\newcommand{\Fun}{\mathit{Fun}}
\newcommand{\Term}{\mathit{Term}}
\newcommand{\Con}{\mathrm{Con}}
\newcommand{\Thm}[1]{\mathit{Thm}(#1)}
\newcommand{\fuelG}{\mathbf{fuel}}
\newcommand{\fuelMu}{\mathbf{fuel}_\mu}
\newcommand{\Lstar}{L^{\!\star}}
\newcommand{\imp}{\supset}

\usepackage{authblk}
\usepackage{doi}

\title{Auto formalisation of Chaitin and of the surprise incompleteness Theorem}
\author{Thierry Coquand}
\affil{Computer Science and Engineering Department,
University of Gothenburg, Sweden,
\href{mailto:coquand@chalmers.se}{coquand@chalmers.se}
}
\date{\today}

\begin{document}
\maketitle

\begin{abstract}
  This is a continuation of a previous report on an experiment in autoformalisation of G\"odel's second incompleteness
  theorem in Agda using Claude. Using the framework built in this experiment, Claude could ``auto-formalise'' Chaitin's proof of the first
  incompleteness theorem and then the Kritchman-Raz surprise examination paradox version of the second incompleteness.

  As the first experiment, the project provides a case study of the strengths
  and limitations of current large language models in mathematics. Since Chaitin's proof involves coding programs, Claude had to represent
  code as ternary string and could build autonomously a parser and a continuation stack evaluation machine. The fact that we can simulate
  computations as expected is not completely trivial and we suggested a Gandy/Howard majorisation argument, that Claude had no problem to
  follow.

  The resulting formalisation clarifies a number of details left implicit in the original
  presentation and provides a fully machine-checked proof of these arguments for Church's Basic Recursive Arithmetic.
\end{abstract}

\section*{Introduction}

This is a continuation of a previous report \cite{arGII}
on an experiment in autoformalisation of G\"odel's second incompleteness
  theorem in Agda using Claude. Using the framework built in this experiment, Claude could ``auto-formalise'' Chaitin's proof of the first
  incompleteness theorem \cite{Chaitin71,Chaitin74}
  and then the Kritchman-Raz surprise examination paradox version of the second incompleteness \cite{KritchmanRaz}.

  We did not write a single line of Agda and it was using a ``spartan'' version, without any tactics, and library. We just want first
  to point out some subtle point about the arguments, that have to be formulated in Skolem arithmetic and on some metatheoretic remark about
  the significance of the proof for constructive mathematics. We then end with a remark on the form of the ``surprise'' argument, which
  as we formulated is actually closer to the sorites/heap of sand paradox than the surprise examination paradox. We let then
  Claude document more in details the formalisation.

  \section{How to encode computation in Basic Recursive Arithmetic}

  Chaitin's proof needs an internal representation of computation. This is subtle since we {\em cannot} define internally an evaluation
  function: it will look like Ackermann function and is not primitive recursive arithmetic. Instead Claude wrote {\em autonomously} and
  without any suggestion from my part a continuation stack small step evaluation machine. The problem is then to show that for a given
  function $f$, is we run the machine on $code(f)$ on an argument $x$, we can find $n$ big enough such that, after $n$ steps, the machine
  halts with the value $f(x)$.

  This is counter-intuitive since there is on the other hand no eval function. Claude was stuck in producing $n$ as a function of $f$. We
  suggested to use the Gandy/Howard method, which produces {\em internally} the number of evaluation steps for computation. With {\em only}
  this suggestion, Claude was able to produce $n$ as an internal function, and then to finish the proof by induction on how $f$ is built.

  \section{Potential interest for constructive mathematics}

  In constructive mathematics, the primary notion is the notion of {\em total} function: a function is ``by nature'' total. It is not so easy
  to find actual examples where the notion of {\em partial} function can play a role. In Chaitin's proof, it plays a crucial role (and this
  in the constructive framework of Binary Recursive Arithmetic). At some point in the argument, we have to exhibit a program which has to be
  {\em short} w.r.t. a certain threshold, and for producing this short program, it is essential to use a looping program, which may be a priori
  partial: we enumerate all possible derivable formuale, $th(0),~th(1),~\dots$ until we find one of a required form. This program is ensured to
  terminate when one uses it, but, in this is the key, the length of the computation is not taken into account in its size.
  (This is a size manifestation of Blum's speed-up phenomenon~\cite{Blum67}: allowing a partial recursive program that is kept total only by an {\em external} proof of termination --- rather than one that must internally certify its own halting --- can shorten the program unboundedly, since program size and running time are independent complexity measures.)

  \section{The statement of Chaitin's result}

  In the setting of Skolem arithmetic, Chaitin's result gets a computational interpretation. It states the existence of a function $g$ such
  that
  $$th(x) = code(K(r)>L*)\Rightarrow th(g(x)) = code(0=1)$$
  for a certain threshold $L*$ which is a concrete, computable,  natural number.
  Here, as usual $K(r)>L*$ means that $r$ cannot be named by a code $p$, such that $log_3(p)\leqslant L*$.
  The theorem is not merely that $K(r)>L*$  is unprovable. It produces an explicit proof-transformer.

  This is somewhat counter-intuitive since there should clearly be numbers that cannot have a short description. This is precisely this which
  is used by Kritchman and Raz \cite{KritchmanRaz} to provide a proof of the second incompleteness theorem.

  \section{The proof of the second incompleteness}

  Our argument for the second incompleteness follows \cite{KritchmanRaz}, but it does not proceed by counting.
  If $N$ is big enough, we know, by a pigeon-hole
  priciple, that we cannot have all numbers in $[0,N]$ having short description. This is $S(0)$. It is then not possible that all numbers
  in $[1,N]$ have short description. The intuitive reason is that, otherwise, we could prove that $0$ has no short description, and then
  we can use Chaitin's Theorem to produce a contradiction. (The exact argument is detailed below.) This is $S(1)$. We can then prove in this
  way by external induction on $r$, the statement $S(r)$ that it cannot be the case that all numbers in $[r,N]$ have a short description.
  We then obtain $S(N+1)$, which is a contradiction.

  We think this argument can also be seen as a variation of the sorites/heap of sand: $[0,N]$ is the heap of sand. If we take away $0$, it
  stays a heap of sand, this is $S(1)$. And if we proceed in this way, we get $S(2),S(3),\dots$ until we have taken away all grains. What is
  striking is that $N$, like $L*$ is a concrete, computable number.

  \section{Conclusion}

  We think that present coding LLM, such as Claude, are now perfect tools to help to understand all details of a given mathematical paper.
  As they are now, they require constant supervisions, which is fine for small examples such as the ones we have analysed. Finally, as
  the previous experiment, the language of type theory was perfect for expressing this kind of concrete mathematics, manipulating
  evaluation machine and coding. The proofs are then directly represented, without any need of tactics.
  We do not need at this level function extensionality and quotient types.
  One key research question may be the design of formal systems where the same can be done for parts of mathematics
  dealing with non-combinatorial object.

{\bf Authorship note.}
Consistent with the autoformalisation theme of this project, the rest of this text, Sections 5–11, were generated by Claude.
(I did not write a single line there.)
The mathematical content, references, and final text were reviewed by the author.

\newpage

\section{The statement}

We work entirely inside the system $T$ of Basic Recursive Arithmetic in
the formulation of Guard, with Shoenfield's notation; the axioms, the
numerals $\mathbf{k}_n = \mathbf{s}^n(\mathbf{O})$, the encoding
$\code{\cdot}$, the verifier $\thmT \in \Fun_1$ (with
$\Thm{\thmT(\code{d}) = \code{A}}$ whenever $d$ codes a derivation of
$A$), the numeral functor $\num \in \Fun_1$ (with
$\Thm{\num(\mathbf{k}_n) = \code{\mathbf{k}_n}}$) and the
derivability-internalising Theorems~12 and~13 are all as set up in the
companion paper~\cite{arGII}, to which we refer for every detail of the
system. We write $\Thm{A}$ for ``$A$ is a
theorem of $T$''.

As in~\cite{arGII}, the consistency of $T$ is the \emph{open} formula
\[
  \Con_T \;\equiv\; \neg\,(\thmT(\mathbf{x}_0) = \code{\mathbf{O}=\mathbf{s}\,\mathbf{O}}),
\]
with the single free variable $\mathbf{x}_0$. Since T has no object
quantifiers, a theorem is an open formula read as implicitly universally
quantified over its free variables, so $\Con_T$ says: \emph{for every
term $\mathbf{x}_0$, $\mathbf{x}_0$ does not code a derivation of}
$\mathbf{O}=\mathbf{s}\,\mathbf{O}$. The theorem we prove is the same
meta-implication that the Guard-route proof establishes,
\[
  \boxed{\;\Thm{\Con_T} \;\Longrightarrow\; \Thm{\mathbf{O}=\mathbf{s}\,\mathbf{O}},\;}
\]
i.e.\ if $T$ proves its own consistency then $T$ is inconsistent. The
arrow $\Longrightarrow$ is a meta-level implication between derivability
judgements, realised in Agda as a total function transforming any proof
of $\Con_T$ into a proof of $\mathbf{O}=\mathbf{s}\,\mathbf{O}$.
The consistency hypothesis is consumed exactly once per step of the
descent of \S\ref{sec:sorites}, and is the \emph{only} assumption: every
other ingredient is constructed.

\section{Chaitin's first incompleteness theorem, internalised}
\label{sec:chaitin}

Chaitin's proof of the first incompleteness theorem~\cite{Chaitin71,Chaitin74}
replaces the liar sentence by Berry's paradox and Kolmogorov complexity.
We need this argument not as a metatheorem but as a single T
implication, with the underlying universal machine realised as an honest
object-level program. This section describes that realisation in
ordinary computational terms; the detailed Agda construction is part of
the development cited in \S\ref{sec:formal}.

\subsection{Programs as ternary numbers}

A \emph{program} is simply a natural number. Read a number $p$ in base
$3$; its digits, drawn from the three-symbol alphabet $\{1,2,3\}$, form a
string, and that string is fed to a fixed parser $\parse$ that turns it
into a syntax tree --- an expression of the object language. The
\emph{length} $|p|$ of the program is the number of base-$3$ digits, so
\[
  |p| \le n \quad\Longleftrightarrow\quad p < 3^{\,n+1}.
\]
Thus the enumeration of programs is the \emph{identity}: the $k$-th
program is the number $k$, and the finite set of programs of length at
most $\Lstar$ is exactly the initial segment
\[
  \{\,p : p < N\,\}, \qquad N := 3^{\,\Lstar+1}.
\]
This is the single simplification that makes the whole argument
feasible: there is no separate enumeration to verify --- ``the set of
short programs'' \emph{is} an interval of numbers, and quantification
over it is bounded quantification over $\{p<N\}$.

\subsection{The universal interpreter as a CK machine}

The interpreter $\evalU$ that runs a parsed program is realised as a small
abstract machine of the Landin CK kind. A \emph{configuration} is a pair
\[
  \langle\, c,\; \kappa \,\rangle,
\]
where $c$ is the expression currently under evaluation (the
\emph{control}) and $\kappa$ is a \emph{continuation stack} recording the
contexts still to be returned to. A one-step transition function $\stepU$
fires a single reduction rule: it either decomposes $c$, pushing the
deferred work onto $\kappa$, or, when $c$ has reached a value, pops the
top frame of $\kappa$ and plugs the value into it. Running the program
for $l$ steps is the $l$-fold iteration $\stepU^{\,l}$ started from the
initial configuration; the machine halts when the control is a value and
the stack is empty, and the output is read off that final configuration.
Around this evaluator we wrap one more layer, a bounded
$\mu$-search loop, so that the outer behaviour of $\evalU$ matches the
informal recipe ``enumerate $\thmT(0), \thmT(1), \dots$ until a fitting
proof is found'' used by the diagonal program below.

For a program $p$ we write
\[
  \defp_p(l, u) \;\equiv\; \text{``running $p$ for $l$ steps halts with output $u+1$''}
\]
(the $+1$ separates ``halted with value $u$'' from ``not yet halted'',
coded by $0$); $\defp_p \in \Fun_2$ for each $p$. Folding the bounded
disjunction of the $\defp_p$ over the finite set $\{p<N\}$ gives a single
object function $\CK \in \Fun_2$ with
\[
  \CK(u, x) = \mathbf{O}
  \;\Longleftrightarrow\;
  \text{some program $p<N$, run for $x$ steps, outputs $u+1$,}
\]
i.e.\ $\CK(u,x)=\mathbf{O}$ expresses $\Kc(u)\le\Lstar$ at run-length
$x$, and the Kolmogorov bound
\[
  \Kc(u) > \Lstar \quad\text{is expressed by}\quad \CK(u,x)\neq\mathbf{O}
\]
with $x$ a free run-length variable.

\subsection{Term-valued fuel: a Gandy--Howard majorant}
\label{sec:fuel}

The delicate point is not the diagonal argument --- which is a faithful
rendering of Chaitin's --- but the fact that the interpreter must be run
\emph{inside} a $\thmT$-fact. Since $\thmT$, $\evalU$ and $\stepU$ all
take object terms, the number of steps appearing in such a fact must
itself be an object \emph{term}, not a meta-level natural; and
Theorem~13 of~\cite{arGII} internalises only statements about T terms.
A completeness statement of the naive form ``for every numeral $\bar n$
large enough, $\stepU^{\,\bar n}$ on the program halts'' is therefore
useless: it is quantified over meta-naturals and cannot be fed into an
internal derivation. What is needed instead is a completeness statement
with a closed \emph{combinator} in the fuel slot,
\[
  (x:\Term)\to\;\mathit{Thm}\,
     \bigl(\,\stepU^{\,\fuelG(f)(x)}(\text{init}(f,x)) = \text{halt}(f\cdot x)\,\bigr),
\]
where $\fuelG(f)\in\Fun_1$ is built once and for all from $f$.

Proving such a statement by induction on the combinator tree of $f$
stalls at the primitive-recursion node, where one would need numerical
control over the running time of the recursive calls --- information not
available at the derivation level. The resolution adapts the classical
\emph{majorisation} method of Howard and Gandy~\cite{Howard73,Gandy80}:
one defines, by primitive recursion mirroring the structure of $f$, a
closed combinator $\fuelG(f)$ that provably \emph{majorises} the actual
number of steps any run of $f$ consumes at every term input. The
induction then carries the stable shape ``the run completes at
\emph{any} fuel $\ge \fuelG(f)(x)$'', which descends through the
recursion node because $\fuelG$ is itself recursive in step with $f$. The
$\mu$-search loop needs the same idea one level higher, with a majorant
$\fuelMu$ stacked on top of $\fuelG$ by a fuel-composition lemma. The
closed term that finally occupies the fuel slot is
$\fuelMu(\cdots) + \fuelG(\cdots)$. Without this majorisation there would
simply be no term to put there.

\subsection{The diagonal program and the internal implication}

Chaitin's short program $g_{\Lstar}$ enumerates $\thmT(0), \thmT(1),
\dots$ until it meets the code of a statement of the form
$\Kc(?)>\Lstar$, and then outputs the witnessed number $?$. Its size is a
constant plus the number of digits needed to write $\Lstar$, that is
$c + \log\Lstar$, which is below $\Lstar$ once $\Lstar$ is large. The
program $g_{\Lstar}$ is obtained as a genuine one-variable object
combinator $G\in\Fun_1$ by bracket abstraction --- an honest T program,
not a meta-level term transformer --- so that $G\cdot w$ is literally the
constructed code, as a \emph{proved} $T$-equation. Let $\mathbf{out}(w)$
be the object function that reads the witness off $w$. The internalised
Chaitin theorem is the single T implication
\[
  \mathit{Thm}\Bigl(\;\bigl(\thmT(w) = \code{\,\Kc(\mathbf{out}(w))>\Lstar\,}\bigr)
        \;\imp\;
        \bigl(\thmT(G\cdot w) = \code{\mathbf{O}=\mathbf{s}\,\mathbf{O}}\bigr)\;\Bigr),
  \tag{C}
\]
valid for \emph{every} term $w$, with no consistency premise. In words:
$T$ proves that \emph{if} $\thmT(w)$ is the incompressibility statement
about $w$'s own read-back, \emph{then} the code $G\cdot w$ is a $T$-proof
of $0=1$. The deduction lives entirely inside the object logic as an
implication; the run of the interpreter on $g_{\Lstar}$, with the
Term-fuel of \S\ref{sec:fuel}, is what closes the positive leg of~(C).
This is the only place where the evaluator machinery is needed; the
descent of the next section uses~(C) purely as a black box.

\begin{remark}[Comparison with Kikuchi's form of Chaitin's theorem]
The form of Chaitin's incompleteness theorem used by Kritchman and
Raz~\cite{KritchmanRaz} is Kikuchi's~\cite{Kikuchi97}, namely the
meta-implication
\[
  \Con_T \;\Longrightarrow\; \forall x\,\neg\,\mathrm{Pr}_T(\code{\,\Kc(x)>\Lstar\,}),
\]
``if $T$ is consistent then $T$ does not prove $\Kc(x)>\Lstar$ for any
$x$'', where $\mathrm{Pr}_T$ is the $\Sigma_1$ provability predicate and
$\Kc$ is the Kolmogorov-complexity function symbol. Our implication~(C)
is sharper in three respects. \emph{First}, it carries no consistency
premise: it isolates the pure diagonal mechanism as an object
implication, and the single appeal to $\Con_T$ is deferred to Step~6 of
the descent. \emph{Second}, it is positive and constructive rather than a
negated provability statement: instead of asserting that $\Kc(x)>\Lstar$
is unprovable, it exhibits an explicit object-level proof-transformer
$G\in\Fun_1$ turning any code of a proof of $\Kc(\mathbf{out}(w))>\Lstar$
into a code of a proof of $\mathbf{O}=\mathbf{s}\,\mathbf{O}$.
\emph{Third}, it is stated with the honest verifier $\thmT$ and the
explicit bounded universal machine of \S\ref{sec:chaitin}, with
$\Kc(x)>\Lstar$ the concrete open $\Pi_1$ formula $\CK(x,\cdot)\neq\mathbf{O}$
over $\{p<N\}$, $N=3^{\Lstar+1}$ --- in particular it uses neither object
quantifiers, nor the abstract $\mathrm{Pr}_T$, nor $\Sigma_1$-completeness,
none of which is available in $T$. Kikuchi's economy --- for instance,
that the only ``monotonicity'' his argument needs is the trivial
$z\le y \wedge y\le \Kc(x) \rightarrow z\le \Kc(x)$ --- stems precisely
from working with $\Sigma_1$ formulas and $\mathrm{Pr}_T$, which bury the
program execution, and its run-length witness, inside the $\Sigma_1$
definition of $\Kc$; lacking object quantifiers, $T$ must instead expose
the run-length as a free fuel variable and pay for it with the genuine
run-monotonicity of Step~1 below.
\end{remark}

\section{The sorites descent}
\label{sec:sorites}

Chaitin's theorem says that, for a large enough $\Lstar$, no statement
$\Kc(x)>\Lstar$ is provable in a consistent $T$, even though for each
$x$ exactly one of $\Kc(x)\le\Lstar$, $\Kc(x)>\Lstar$ holds and the
former, when true, is provable (by exhibiting and running the short
program). Kritchman and Raz~\cite{KritchmanRaz} observed that this
tension can be wound into the second incompleteness theorem by an
argument modelled on the surprise-examination paradox.

\subsection{Induction on the stages, not counting; a sorites}

Kritchman and Raz~\cite{KritchmanRaz} argue by \emph{counting}: writing
$m = \#\{\,u\in[0,N] : \Kc(u)>\Lstar\,\}$, they prove $\Thm{m\ge i}$ for
$i = 1, 2, \dots$ inside $T$, and since $m \le N+1$ this induction must
eventually clash. We do not count. Instead, fixing $\Lstar$ as in
\S\ref{sec:chaitin} and $N = 3^{\Lstar+1}$, we prove directly, by
external induction on $r$, the open stage statement $S(r)$ of
\S\ref{sec:sixsteps} --- ``not all of the days $[r,N]$ can be
simultaneously described by short programs'' --- with base case $S(0)$ the
pigeonhole below and inductive step $S(r)\to S(r+1)$ the six-step block,
so that the chain $S(0)\to S(1)\to\dots\to S(N)$ closes with
$\Thm{\mathbf{O}=\mathbf{s}\,\mathbf{O}}$.

This has the shape of the \emph{sorites}, or heap-of-sand, paradox, rather
than a vicious self-referential circle. Picture the interval $[0,N]$ as a
heap of sand: each step of the induction removes one grain and is, taken
on its own, an entirely sound $T$ derivation --- one internal application
of the Chaitin implication~(C) reflected through a single use of
consistency --- so no individual grain-removal is the culprit, yet after
$N+1$ removals the heap is gone and $T$ has proved
$\mathbf{O}=\mathbf{s}\,\mathbf{O}$. The contradiction lives only in the
iteration, exactly as in the heap paradox, and it is the second
incompleteness theorem --- the impossibility of $T$ proving its own
consistency --- that explains why the seemingly innocuous step cannot in
fact be taken all the way (cf.~\cite{KritchmanRaz}: the missing premise is
the consistency of $T$ itself).

\subsection{The formal realisation: six steps}
\label{sec:sixsteps}

In the formalisation the count $m$ is carried by an open formula
$S(r)$, indexed by a stage $r \in [0,N]$ and built as a negated big
conjunction over ``days''. For programs $p_r,\dots,p_N$ (each of length
$\le\Lstar$, ranging over $\{p<N\}$) write
\[
  \Kc(\mathbf{x}_0; p_r,\dots,p_N) \;\equiv\;
    \bigwedge_{d=r}^{N} \defp_{p_d}(\mathbf{x}_0, d),
\]
``each program $p_d$ describes the day $d$, at the common run-length
$\mathbf{x}_0$''. The stage formula is
\[
  S(r) \;\equiv\;
    \Thm{\neg\,\Kc(\mathbf{x}_0; p_r,\dots,p_N)}
    \qquad\text{(for every choice of $p_r$).}
\]
The descent climbs $S(0) \to S(1) \to \dots \to S(N)$, one grain at a
time, and the inductive step $S(r)\to S(r+1)$ is the six-step block that
realises Kritchman--Raz's induction step; it follows the high-level note
\texttt{clos} of the development and reuses the implication~(C) as a
black box.

\begin{description}
\item[Step 1 (run-monotonicity and the finite set).]
From $S(r)$ one derives, at an \emph{independent} run-length
$\mathbf{x}_1$, the implication
\[
  \Kc(\mathbf{x}_0; p_{r+1},\dots,p_N) \imp \neg\defp_{p_r}(\mathbf{x}_1,r),
\]
the two fuels
$\mathbf{x}_0,\mathbf{x}_1$ being aligned to a common bound by
monotonicity of the interpreter in its step count (running longer never
unsays a halt). Folding the bounded disjunction over the finite set
$\{p<N\}$ rewrites the consequent as $\Kc(r) > \Lstar$, giving
\[
  \mathit{Thm}\bigl(\,\Kc(\mathbf{x}_0; p_{r+1},\dots,p_N) \imp (\Kc(r)>\Lstar)\,\bigr).
\]

\item[Step 2 (encode and substitute the numeral).]
By Theorem~13 of~\cite{arGII} the Step-1 implication is internalised: one
obtains a term $f$ with
\[
  \Thm{\thmT(f\cdot\mathbf{x}_0)
   = \code{\,\Kc(\num\,\mathbf{x}_0; p_{r+1},\dots,p_N) \imp \Kc(r)>\Lstar\,}},
\]
the subject slot $\mathbf{x}_0$ being replaced inside the code by its
numeral $\num\,\mathbf{x}_0$ (the consequent has no free $\mathbf{x}_0$
and is left inert).

\item[Step 3 (internal modus ponens).]
An encoded, Carneiro-lifted modus ponens~\cite{arGII} chains Step~2 with
a $\thmT$-fact for the antecedent.

\item[Step 4 (Theorem~13 for the conjunction).]
Writing the big conjunction as $K_r(\mathbf{x}_0)=\mathbf{O}$, the
$\Sigma_1$-completeness direction
\[
  \begin{aligned}
  \bigl(K_r(\mathbf{x}_0)=\mathbf{O}\bigr) &\imp \thmT(D\,\mathbf{x}_0) = {}\\
  &\quad \code{\,\Kc(\num\,\mathbf{x}_0; p_{r+1},\dots,p_N)\,}
  \end{aligned}
\]
supplies that antecedent fact (Kritchman--Raz's equation~(2)).

\item[Step 5 (Chaitin).]
The internal Chaitin implication~(C) is applied:
\[
  \mathit{Thm}\bigl(\,(\thmT(w)=\code{\Kc(r)>\Lstar})
   \imp (\thmT(G\cdot w)=\code{\mathbf{O}=\mathbf{s}\,\mathbf{O}})\,\bigr).
\]
Composing Steps~3--5 yields
\[
  \mathit{Thm}\bigl(\,K_r(\mathbf{x}_0)=\mathbf{O}
   \imp (\thmT(h\cdot\mathbf{x}_0)=\code{\mathbf{O}=\mathbf{s}\,\mathbf{O}})\,\bigr).
\]

\item[Step 6 (consistency).]
Finally $\Con_T$ is instantiated at $h\cdot\mathbf{x}_0$ and, by
contraposition, refutes the consequent --- so the antecedent fails:
$\Thm{\neg\,\Kc(\mathbf{x}_0; p_{r+1},\dots,p_N)}$, which is $S(r+1)$.
This is the single use of consistency per grain (Kritchman--Raz's
equation~(1)).
\end{description}

\paragraph{Base case.}
$S(0)$ is hypothesis-free and combinatorial. There are $N+1$ days but
only $N$ program slots, so any assignment of a program to each day forces
a collision $p_i = p_j$ with $i\neq j$; the two day-conjuncts then assert
that one program, at the same run-length, outputs both $\mathbf{s}^i$ and
$\mathbf{s}^j$, which is impossible. Hence $\neg\Kc(\mathbf{x}_0;p_0,\dots,p_N)$.

\paragraph{Final clash.}
At the top stage $r=N$ the remaining conjunction is empty, hence
provably true, so the Step-1 implication fires \emph{unconditionally}:
$T$ proves $\Kc(N)>\Lstar$, the Chaitin diagonal makes its provability
into a proof of $0=1$, and $\Con_T$ refutes it. The heap is empty and
\[
  \Thm{\mathbf{O}=\mathbf{s}\,\mathbf{O}}.
\]
By external induction the chain $S(0)\to\dots\to S(N)$ closes the proof.
This induction is not informal: it is carried out in the metatheory by
Agda, as a recursion on the meta-natural $r$ that turns the six-step
block into a function
$\Thm{S(r)} \to \Thm{S(r+1)}$ and iterates it from
the base derivation of $S(0)$ up to $S(N)$.

\section{The formalisation}
\label{sec:formal}

The proof is checked in Agda, in the same \texttt{T4} development as the
Guard-route proof of~\cite{arGII}, under the global options
\texttt{-{}-safe -{}-without-K -{}-exact-split}, with no postulates and no
holes. The headline is a single function from a derivation of $\Con_T$ to
a derivation of $\mathbf{O}=\mathbf{s}\,\mathbf{O}$; $\Con_T$ is the only
hypothesis, and it is the same open consistency formula as
in~\cite{arGII}.

The one genuinely non-classical piece is the Gandy--Howard majorant of
\S\ref{sec:fuel}. The diagonalisation is Chaitin's, but the closed object
term occupying the fuel slot of the internal interpreter run has no
counterpart in the informal argument, and its construction (by recursion
mirroring the program, with a substituted-motive induction through the
recursion node) is what lets evaluator completeness be \emph{stated and
proved} inside T. It is worth noting, finally, that Guard's
formulation of Basic Recursive Arithmetic --- syntax-first, with the
verifier $\thmT$ and the numeral functor $\num$ as honest object
functions --- proved well suited to expressing all of these results,
including the internal Chaitin implication and the surprise descent.

The development, including both proofs of the second incompleteness
theorem, is available at
\url{https://github.com/coquand/agda-godel-tree}.

\end{document}